# Novel magnetic excitations beyond the single- and double-magnons

Hebatalla Elnaggar,[1,2]* Abhishek Nag,[3] Maurits W. Haverkort,[4] Ke-jin Zhou,[3] Frank de Groot[1]*

[1]Debye Institute for Nanomaterials Science, Utrecht University, 3584 CA Utrecht, The Netherlands.
[2]Institute of Mineralogy, Physics of Materials and Cosmochemistry, Sorbonne University, 4 Place Jussieu, 75005 Paris, France.
[3]Diamond Light Source, Harwell Campus, Didcot OX11 0DE, United Kingdom.
[4]Heidelberg University, Philosophenweg 19, 69120 Heidelberg, Germany.

**Abstract**

Conventional wisdom suggests that one photon that carries one unit of angular momentum can change the spin angular momentum of a magnetic system with one unit ($\Delta M_s = \pm 1$) at most. This would imply that a two-photon scattering process can manipulate the spin angular momentum of the magnetic system with a maximum of two units. Here we examine the fundamental limit of the photon-driven transport of angular momentum by studying the magnon spectrum of $\alpha$-Fe$_2$O$_3$ using resonant inelastic x-ray scattering. We discovered a cascade of higher-rank magnons carrying double, triple, quadruple, and quintuple the spin angular momentum of a single-magnon. Guided by theoretical calculations, we reveal how a two-photons scattering process can create exotic higher-rank magnons and the relevance of these quasiparticles for magnon-based applications.

**Introduction**

Understanding how to control the spin degree of freedom is a cornerstone for several hot topics of contemporary magnetism research, including ultrafast magnetism and magnonics. The main idea behind magnonics is to use the elementary magnetic excitations (magnons) for information transfer and processing. Magnons are bosonic quasiparticles and are the quanta of magnetic oscillations of systems with periodically ordered magnetic moments (1). A magnon is classically depicted as a phase-coherent precession of microscopic vectors of magnetization in a magnetic medium. When a magnon propagates through a magnetic medium, no electrical charge transport is involved and hence no electrical losses take place. This is the key advantage of using magnons as information carriers. The energy of magnons range typically in the terahertz range (in the order of 1 to 25 THz, i.e. 5 to 100 meV). The magnon frequency has an important impact on the performance of magnon-based devices because the larger the excitation frequency, the "faster" are the magnons. This means that the use of high-frequency (terahertz) magnons would provide a great opportunity for the design of ultrafast devices (2). Antiferromagnets represent an appealing playground for the search for new channels of high frequency, long-lived magnons paving the way towards ultrafast magnon-based devices (3,4).

Collective excitations such as magnons can be effectively measured using *2p3d* resonant inelastic x-ray scattering (5). This realization has been the main motivation behind the development of ultra-high resolution resonant inelastic x-ray scattering (RIXS) beamlines, with the goal to study the spin dynamics of (pseudo)spin S = ½ materials such as cuprates and iridates (6-9). Spin half systems represent a special case as only transitions from $M_s$ = -½ to $M_s$ = ½ are allowed on a single atomic site. These magnons propagate a change of one unit of angular momentum and are similar to the magnons observed by other techniques such as inelastic neutron scattering and Raman scattering. We will refer to these magnons as conventional single-magnons. As RIXS involves two-photons, each carrying one unit of angular momentum, the angular momentum conservation suggests that the RIXS process can change the spin angular momentum of the system with a maximum of two units. Our recent RIXS measurements on NiO confirmed this where we observed single ($\Delta M_s = 1$) and double-magnons ($\Delta M_s = 2$) within a single $Ni^{2+}$ magnetic site (10). We point out that double-magnons are different from bimagnons observed in cuprates. A double-magnon is a $\Delta M_s = 2$ transition at a single site, while a bimagnon is composed of two single-magnons being excited on two different magnetic sites, one changing the spin projection with +1 (i.e., $\Delta M_s = 1$) and the other with -1 (i.e., $\Delta M_s = -1$) giving rise to a combined $\Delta M_s = 0$ transition (11).

While it is clear that for high spin $Ni^{2+}$ ions in NiO possessing two unpaired $3d$ electrons that only two spins can change their angular momenta (i.e. excitations between $M_s = 1, 0, -1$), the situation is more complicated for a high spin $3d^5$ ion in a magnetic system. In this case there are conceptually five spins that could be locally reversed resulting in magnons carrying a change of angular momentum up to 5 units (i.e. excitations between $M_s = 5/2, 3/2, ½, -½, -3/2, 5/2$). This raises the fundamental question: Is it possible to excite higher-rank magnons that carry double, triple, quadruple, and quintuple the spin angular momentum of a single-magnon using a two-photon process such as RIXS?

**Results**

We provide the first experimental data capable of answering this question by measuring the low-energy magnon spectrum of α-$Fe_2O_3$ single crystal at the ultra-high resolution I21 RIXS setup ($\Delta E = 32$ meV) at Diamond Light Source (12). The antiferromagnet α-$Fe_2O_3$ is an ideal material to initiate this kind of study because the ground state of $Fe^{3+}$ has the maximum number of unpaired electrons for the $d$ orbitals providing a platform to test the maximum number of possible single-site spin-flip excitations. Furthermore, the $^6A_1$ orbital singlet ground state makes a clean case to study solely spin excitations without any orbital contribution.

Figure 1(A) shows the hexagonal unit cell of α-$Fe_2O_3$ with the corundum structure. The exchange coupling is dominantly antiferromagnetic with the Néel temperature $T_N$ of ~950 K. In addition to the Néel transition, α-$Fe_2O_3$ exhibits another magnetic transition, referred to as the Morin transition ($T_M$ of ~250 K) where below $T_M$ it is purely antiferromagnetic. We performed our measurements at 13 K ($T < T_M$) in the collinear antiferromagnetic phase. An exemplary $L_3$ x-ray absorption spectrum (XAS) is shown in Figure 1(B) where we find two main peaks (labelled $E_1$ and $E_2$) exhibiting the expected x-ray magnetic linear dichroism signal in line with previous literature (13-15).

The RIXS spectrum measured at $E_1$ is shown in Figure 1(C) and an overview of the RIXS spectra with extended energy-loss range are shown in Figure S1. The elastic line is observed at zero energy transfer where the final state preserves the initial state spin orientation. A cascade of energy transfer peaks can be seen at 100, 200, 300, 400 and 500 meV. The first energy transfer peak can be assigned to a single-magnon excitation. The mean-field exchange interaction in α-$Fe_2O_3$ is ~100 meV as would be expected from its Néel temperature and agrees well with the observation from inelastic neutron scattering experiments where an optical nearly non-dispersing mode is observed at ~100 meV (16). The single-magnon excitation represents a single spin-flip where the change of angular momentum propagated by the quasiparticle is $\Delta M_s = 1$ as depicted in the cartoon in the left of Figure 1(C). The second energy transfer peak appears at double the energy of the single-magnon and can be assigned to a single-site double-magnon excitation ($\Delta M_s = 2$) similar to the double-magnon excitation that we observed in NiO (10).

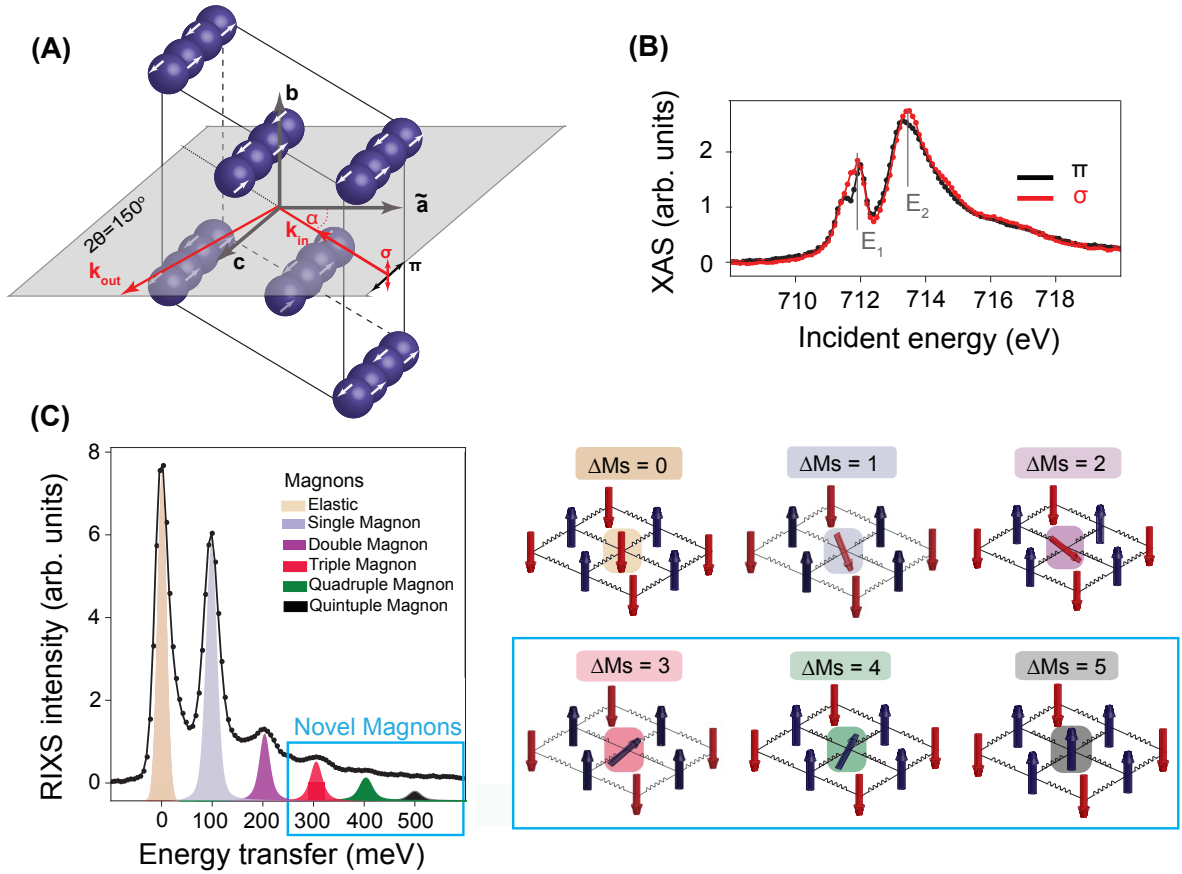

**Figure 1: Crystal structure and x-ray L$_3$-edge measurements in a (0001) α-Fe$_2$O$_3$ single crystal.**
**(A)** Crystal structure of α-Fe$_2$O$_3$ and the scattering geometry used for all measurements presented in this work. k$_{in(out)}$ are the incident and outgoing wave vectors and the scattering angle (2θ) was kept fixed at 150°. The incidence angle is α with α = 90° for normal incidence. The antiferromagnetic order is depicted with white arrows showing the orientation of the magnetic moments. **(B)** Fe L$_3$ XAS measured with π (black) and σ (red) polarization. Two main peaks can be identified and are labelled E$_1$ and E$_2$. **(C)** Right: RIXS spectra measured at E$_1$ (α = 95°, π polarization). The orange shaded Gaussian peak at zero energy transfer corresponds to the elastic peak also having contribution from ΔM$_S$=0 excitation. The five shaded antisymmetric Lorentzian peaks represent the single- (blue), double- (purple), triple- (red), quadrupole- (green) and quintuple- (black) magnon excitations (see Methods for the fitting details). Right: A sketch depicting the excitations observed by RIXS.

The most remarkable feature in our results is the ability of the two-photon RIXS process to excite higher-rank magnons at 3 times, 4 times and potentially a broad 5 times the energy of a single-magnon (see Figure 1(A)). These higher-rank magnons carry these higher multiples of angular momentum. We provide the details of the fitting in the materials and methods and Figures S2 and S3. We measured the angular dependence of the single-, double- and triple-magnons by rotating the sample in the azimuthal direction to decipher the nature of the transitions involved (see Figure 2). Conceptually, one expects that the angular behavior of the higher-order magnons differs from the single-magnon as the angular momentum selections rules are different (ΔM$_s$ = 1, 2, 3 involves a dipolar, quadrupolar and hexapolar spin-flip processes respectively). This is confirmed by comparing the angular behavior in panels (A), (B) and (C) of Figure 2 where the magnitudes of the transitions are reduced approximately with an order of magnitude as we move from single- to double- to triple-magnons in addition to the change of the angular profile. We measured the *q*-dependence of these excitations however they showed nearly no dispersion (see Figure S4). This is not unexpected as the 100 meV optical mode shows negligible dispersion according to inelastic neutron scattering measurements (16).

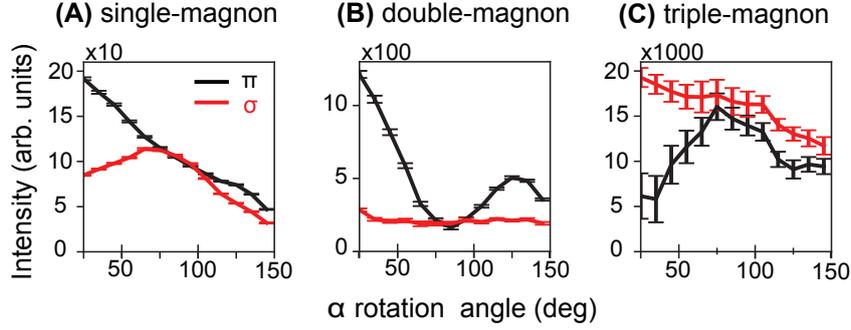

**Figure 2: Angular dependence of the magnons measured with π and σ polarization.** (A) Single-magnon, (B) double-magnon and (C) triple-magnon excitations measured at the incidence energy $E_1$. The angular dependence is measured by rotating the single crystal in the azimuthal direction (α rotation) while the scattering angle was kept fixed at 150°.

**Discussion**

The implication of our experimental observation is that the two-photon RIXS process can exchange five units of angular momenta with a single magnetic-site. This is an unexpected result because the selection rule for a dipole transition in presence of strong spin-orbit coupling is that the change of the total angular momentum ($\Delta M_j$) is equal to 0 or ±1. This means that for a dipole-in (2p → 3d transition), dipole-out (3d → 2p transition) 2p3d RIXS process, possible transition should involve $\Delta M_j$ = 0, ±1, ±2 giving rise to single- and double-magnons. We performed multiplet ligand field theory calculation for $Fe^{3+}$ $L_3$-edge RIXS (see Figure 3(A)) which confirms that only single- and double- magnons are expected to be observed and is in line with our previous work on NiO (10). It is essential to examine the interaction terms of the model Hamiltonian responsible for the single- and double-magnons in an attempt to find the origin behind the higher-ranked magnons in α-$Fe_2O_3$.

$$H = \sum_k f_k F^k + \sum_k g_k G^k + \sum_i l_i \cdot s_i + J_{exch}(n.S) \tag{1}$$

The model Hamiltonian used for the calculation is given by equation (1). The $J_{exch}(n.S)$ term is the mean-field super exchange interaction term that determines the energy of the single-magnon. The spin orbit coupling is given by the $\sum_i l_i \cdot s_i$ term and is responsible for the spin-flip process by mixing the orbital and spin degrees of freedom and enables the observation of single-magnons as detailed in the work of Groot *et. al.* (17). The double-magnon is enabled through the intra-atomic Coulomb exchange given by $\sum_k f_k F^k + \sum_k g_k G^k$. The intra-atomic Coulomb exchange interaction strongly couples the valence and core electrons implying that the spin angular momentum of both the core and valence orbitals are no longer a good quantum number, effectively leading to $\Delta M_s$ = 0, 1, and 2 excitations.

It is inevitable to conclude that the higher-rank magnons have a different origin compared to the single- and double-magnons reported in previous work (10,18) and a mechanism that allows the exchange of higher angular momenta is needed. This apparent contradiction can be reconciled by realizing that the angular momenta of electrons only is not a conserved quantum number in real crystals (19). The crystal lattice can exchange momentum with the electrons providing the extra angular momentum required for higher-rank magnons involving $\Delta M_S > 2$. We performed a full-multiplet ligand field theory calculation for $Fe^{3+}$ $L_3$-edge RIXS including the effect of the crystal lattice using an effective octahedral crystal field potential (see Figure 3(B)). In addition to the single- and double-magnons, triple-, quadruple- and quintuple-magnons are now visible confirming that the crystal field potential is the key factor behind the higher-rank excitations. Here we stress that the higher-rank magnons are generated from a single magnetic-site and the role of the crystal lattice can be considered as a reservoir of angular momentum supplying the extra angular momentum required.

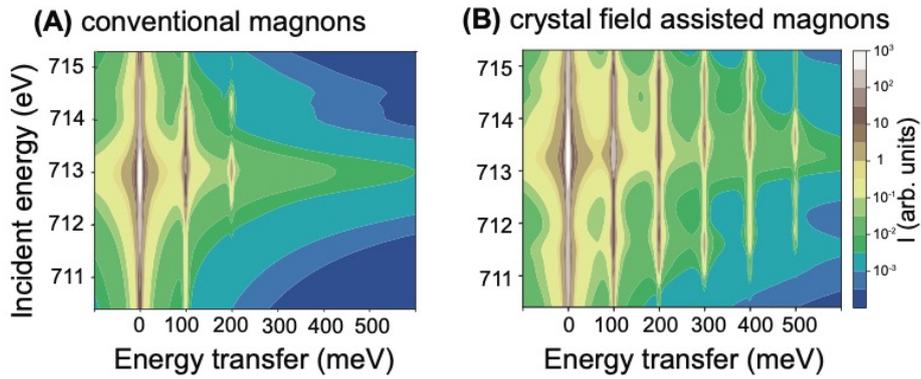

**Figure 3. Full-multiplet ligand field theory calculation for $Fe^{3+}$ $L_3$-edge RIXS in α-$Fe_2O_3$. (A)** For an $Fe^{3+}$ ion according to the Hamiltonian in Equation 1. **(B)** Considering an additional term that includes the crystal field effects. The calculations were performed for linear horizontal incoming beam and unpolarized outgoing beam to correspond to the experimental conditions. The parameters used for the calculations are reported in SI Tab. 1.

To visualise the role of the crystal lattice, we follow the fate of an excitation created by the absorption, for example, of a circular right polarized photon[1]. The initial state can be represented by the vector shown in the upper left corner of Figure 4 which includes the *2p* and *3d* orbitals participating in the RIXS process. The first two numbers shaded in gray represent the occupation of the *2p* spin down (red arrow) and *2p* spin up orbitals (blue arrow) where we take the spin quantization axis to be the $C_4$ axis of the octahedron. The second two numbers shaded in yellow are the *3d* spin down (red arrow) and *3d* spin up orbitals (blue arrow) occupation numbers. The projection of the total orbital angular momentum, $L_Z$, is specified in the subscript of the vector to keep track of the orbital angular momentum of the states. This means that the initial state is given by $(3,3,5,0)_0$. Upon the absorption of the photon, an electron can be excited from the *$2p_{-1}$* to the *$3d_{-2}$* orbital resulting in an intermediate state given by $(3,2,5,1)_{-1}$ (Figure 4(A)). This intermediate state can decay back elastically by emitting a right polarized photon (Figure 4(B1)). Another possible path is shown in Figure 4(B2) where a *$3d_{-2}^↑$* electron scatters of the crystal field potential to a *$3d_2^↑$* orbital and thereby changes its angular momentum by four units. This extra angular momentum provided by the lattice is the key aspect that makes it possible to excite higher-rank magnons. A cascade of *2p* spin-orbit coupling and *2p-3d* exchange interaction are required to transfer this orbital angular momentum to spin angular momentum as illustrated in panels (C) to (I) of Figure 4. The first pair of *2p* spin-orbit coupling and *2p-3d* exchange interaction changes the intermediate state to $(3,2,4,2)_2$. We note that this intermediate state cannot decay to a single-magnon excitation as it cannot reach $L_Z = 0$ final state through a dipole emission. The second pair of *2p* spin-orbit coupling and *2p-3d* exchange interaction changes the intermediate state to $(3,2,3,3)_1$. Now this intermediate state can decay to a double-magnon excitation either after the *2p* spin-orbit coupling step ($2p_{-1}^\downarrow \to 3d_{-2}^\downarrow$ emitting a left polarized photon - not shown in Figure 4 for visual clarity) or after the exchange interaction step emitting a left polarized photon (Figure 4(G1)). A final *2p* spin-orbit coupling step is required to changes the intermediate state to $(2,3,3,3)_0$ which can finally decay to a triple-magnon by emitting a Z polarized photon (Figure 4(H)). Quadrupole and quintuple magnons can be reached by further exchange of angular momentum with the lattice followed by cascades of *2p* spin-orbit coupling and *2p-3d* exchange interaction. This transparent Feynman-diagram representation allows us to derive simple selection rules for this example: (i) it is not possible to excite single-magnons using circular right polarized incoming x-rays, (ii) double-magnons can be selectively observed by detecting the left polarized light out. (iii) triple-magnons can be selectively observed by detecting the Z polarized light out. The Z polarized light can be experimentally detected by placing an extra detector in the vertical plane for example. A full RIXS calculation is shown in Figure S5 and confirms the selection rules derived from Figure 4.

---

[1] We define the polarization of light as: $|R\rangle = \frac{1}{\sqrt{2}}\begin{pmatrix}1\\-i\\0\end{pmatrix}$, $|L\rangle = \frac{1}{\sqrt{2}}\begin{pmatrix}1\\i\\0\end{pmatrix}$ and $|Z\rangle = \begin{pmatrix}0\\0\\1\end{pmatrix}$.

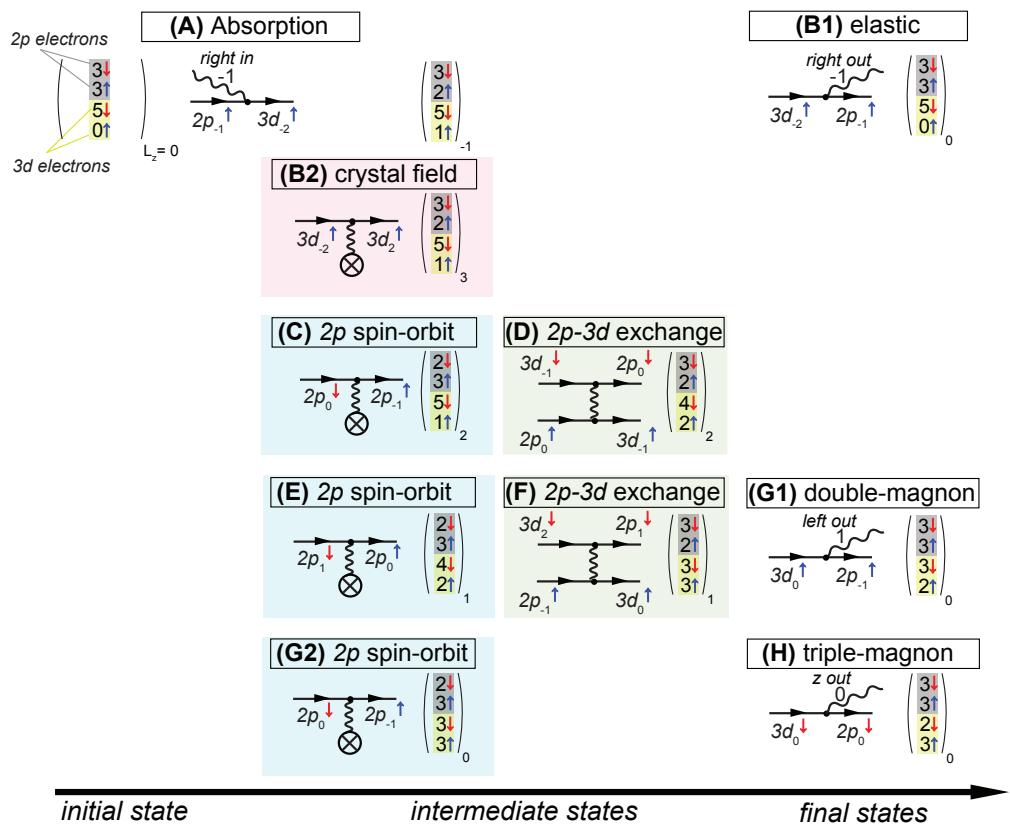

**Figure 4. Schematic representation of the mechanism of higher-rank magnons excitations by *2p3d* RIXS.** The initial state vector is shown in the upper left corner comprising of the *2p* (gray) and *3d* (yellow) orbitals participating in the RIXS process. The spin of the electrons is depicted by the colored arrows (red = down, blue = up). We follow the fate of a *2p→3d* excitation created by the absorption of a right polarized photon (A) up to the triple-magnon decay through a cascade of crystal field interaction, *2p* spin-orbit coupling and *2p-3d* exchange interaction through the steps from **(B)** to **(H)**.

We developed a low-energy effective RIXS operator that describes low-energy magnetic excitations such as magnons in terms of spin operators based on the work of Haverkort (20). The full RIXS cross-section is given by equation (2) where the ground state $|i\rangle$ is excited by a photon described by a dipole transition operator $T_{\epsilon_i}$ to an intermediate state described by the Hamiltonian (H) and decays to all possible final states $|f\rangle$ emitting a photon described by a dipole transition operator $T_{\epsilon_o}$.

$$RIXS(\omega) \propto \sum_f |\langle f|T_{\epsilon_o}^{\dagger} \frac{1}{\omega_i - H + i\Gamma/2} T_{\epsilon_i}|i\rangle|^2 = \sum_f |\langle f|R_{eff}|i\rangle|^2 \quad (2)$$

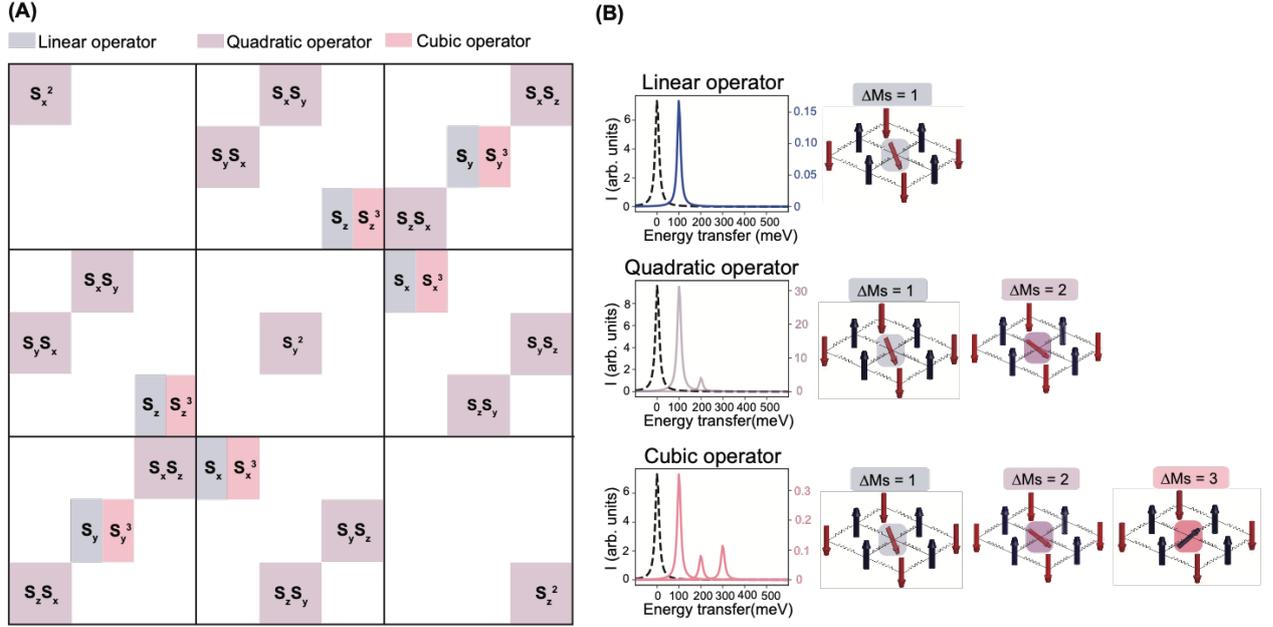

**Figure 5. Matrix elements of the effective RIXS operator ($R_{eff}$) expanded up to the third rank and the RIXS spectrum from individual operators.** (**A**) A summary of the expansion presented in equation 3 grouped in terms of linear (blue), quadratic (purple) and cubic (red) spin operators (S) involved in the RIXS cross-section. (**B**) The RIXS cross-section computed using the three orders of the spin operators.

The effective operator ($R_{eff}$) removes the intermediate state from the equation by expanding the intermediate state Hamiltonian in terms of polynomials of spin operators multiplied by x-ray absorption fundamental spectra. The expression of the expansion to the third order is presented in the materials and methods and is summarized in Figure 5(A). The spin flip processes resulting in magnons can be grouped in order of the spin operator rank: linear (shaded in blue), quadratic (shaded in purple) and cubic (shaded in red) operators. The linear spin operators can generate single-magnons while the quadratic spin operators generate single- and double-magnons and produce the majority of the RIXS intensity (see Figure 5(B)). Finally, the cubic spin operators can generate single-, double- and triple-magnons. The main advantage of this expansion is its simple form that allows one to determine general selection rules depending on the polarization of the incoming and outgoing light. We computed the angular dependence of the single-, double- and triple-magnons based on this expansion in Figure 6. Our calculations capture all essential aspects of the experimental angular dependence where we obtain the correct order of magnitude for the magnons and reproduce the angular dependence well confirming the nature of the higher-rank magnons (compare Figure 2 to Figure 6). The minor deviation of the angular dependence could be a result of saturation affects due to the sample rotation (21).

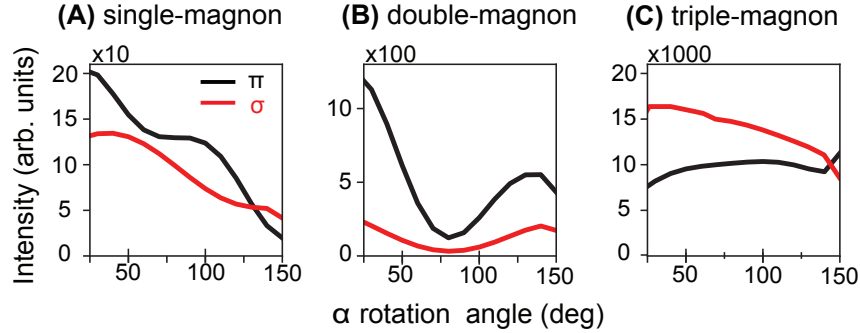

**Fig. 6. Computed angular dependence of the magnons measured with π and σ polarization using the effective operator expansion. (A)** Single-magnon, **(B)** double-magnon and **(C)** triple-magnon excitations at the incidence energy $E_1$.

## Conclusion

The comparison between theory and experiment on α-$Fe_2O_3$ confirms that the cascade of excitations we observed at triple, quadruple, and quintuple the energy of a single-magnon are higher rank-magnons that propagate these higher multiples of spin angular momentum. From a fundamental point of view, the higher-rank magnons can couple differently with the various degrees of freedom of the system providing a unique platform to investigate magnon interactions. From a technological point of view, these excitations have higher energies than that of single-magnons and hence are potentially more thermally robust. We predict that these higher-rank magnons can also be excited using THz-pulses and can be enhanced using magnonic crystals paving the way towards future magnonic devices.

## Materials and Methods

### *Resonant inelastic x-ray scattering measurements*
High-resolution ($\Delta E$= 32 meV), Fe $L_3$-edge resonant inelastic x-ray scattering measurements were done at the I21 beamline of the Diamond Light Source, United Kingdom. Linear horizontally (π) or vertically (σ) polarized x-ray beam was used. The angular dependence was measured by rotating the sample about the b-axis (referred to as α rotation) of a polished α-$Fe_2O_3$ single crystal cooled down to 13 K. The Morin temperature of $Fe_2O_3$ is 220 K. The scattering angle was kept fixed at 150°. The x-ray absorption spectrum (XAS) shown in Figure 1(B) was measured using total electron yield in the same geometry. The pressure in the experimental chamber was maintained below $5\times10^{-10}$ mbar. The zero-energy transfer position and resolution of the RIXS spectra were determined from subsequent measurements of elastic peaks from an adjacent carbon tape.

### *Resonant inelastic x-ray scattering data fitting*
RIXS data were corrected for self-absorption prior to fitting. The elastic line was fitted with an energy resolution limited Gaussian lineshape (orange shade, Figure 1C and SI Figure 1A). The phonon peaks close to 43 meV and 150 meV were fitted with asymmetric Lorentzian functions (shown by gray dashed lines) convoluted with the energy resolution. These peaks are clearly visible in RIXS spectra at $E_2$ peak of XAS (see inset of SI Figure 1B). The five shaded antisymmetric Lorentzian peaks convoluted with the energy resolution represent the single- (blue), double- (purple), triple- (red), quadrupole- (green) and quintuple- (black) magnon excitations (Figure 1C and SI Figure 1A). While the energy positions up to the quadrupole magnon excitation were kept as a free parameter for fitting, the energy position of the quintuple magnon was fixed to (fitted energy position of the triple magnon)*(5/3). See also SI Figure 2 and SI Figure 3 for the low energy fits to the RIXS data for different α at π and σ polarisations, respectively.

*Multiplet ligand-field calculations*

The crystal field multiplet model is an effective model Hamiltonian for the description of all charge conserving excitations of ionic transition metal systems. The crystal field multiplet model is valid for the main peaks of *2p* x-ray absorption and the low-energy RIXS excitations of ionic transition metal ions, because the *2p3d* x-ray absorption and the *3d2p* x-ray emission are neutral, self-screened, transitions, which implies that screening channels such as ligand-metal charge transfer can be approximated by renormalized parameters. We used the quantum many-body program Quanty (22) to simulate Fe XAS and *2p3d* RIXS in α-$Fe_2O_3$. The Hamiltonian used for the calculations consists of the following terms: (ii) crystal field potential, (iii) spin−orbit coupling, and (iv) exchange interaction. The *d–d* (*p–d*) multipole part of the Coulomb interaction was scaled to 70% (80%) of the Hartree−Fock values of the Slater integral. The general parameters used for the simulations are in agreement with previous studies of α-$Fe_2O_3$ $L_{2,3}$ edges.

*Expression of the effective RIXS operator*

The effective operator can be expressed by equation (3) as shown by Haverkort (20). Here $\epsilon_{in(out)}$ is the incoming (outgoing) polarization of the photons. $F_{\{x,y,z\}}$ is the conductivity tensor describing the full magneto-optical response function of the system depending on the local magnetization direction given by $\{x, y, z\}$.

$$R_{eff} = -Im[\epsilon_{in}^* \cdot F_{\{x,y,z\}} \cdot \epsilon_{out}] \tag{3}$$

The general form of the conductivity tensor can be expressed as a sum of linear independent spectra multiplied by functions depending on the local magnetization direction as given in equation (4).

$$F_{\{x,y,z\}} = \sum_{k=0}^{\infty} \sum_{m=-k}^{k} \begin{pmatrix} F_{xx}^{k,m} & F_{xy}^{k,m} & F_{xz}^{k,m} \\ F_{yx}^{k,m} & F_{yy}^{k,m} & F_{yz}^{k,m} \\ F_{zx}^{k,m} & F_{zy}^{k,m} & F_{zz}^{k,m} \end{pmatrix} Y_{k,m}(\theta, \phi) \tag{4}$$

Here $\theta$ and $\phi$ define the direction of the local moment with $\theta$ being the polar angle, and $\phi$ being the azimuthal angle. $Y_{k,m}(\theta, \phi)$ is a spherical harmonic function and $F_{ij}^{k,m}$ is the i, j component of the conductivity tensor on a basis of linear polarized light in the coordinate system of the crystal. In symmetries lower than spherical, this expansion on spherical harmonics does not truncate at finite $k$ and the angular momentum of the electrons only is not a conserved quantum number in the crystals. We have shown in our previous work that including terms up to $k = 3$ is sufficient to describe $Fe^{3+}$ ions in octahedral crystal field (23). The expression is given in equation (5) and involves terms up to the 3$^{rd}$ order in spin leading to single, double and triple spin flip processes.

$$F_{\{x,y,z\}} =$$

$$\begin{pmatrix} F_{a1g}^0 + 2F_{eg}^2(S_x^2 - \frac{1}{3}S^2) & F_{t2g}^2(S_xS_y + S_yS_x) - F_{t1u}^1 S_z - F_{t1u}^3(-\frac{3S_z}{5} + S_z^3) & F_{t1u}^1 S_y + F_{t1u}^3(-\frac{3S_y}{5} + S_y^3) + F_{t2g}^2(S_xS_z + S_zS_x) \\ F_{t2g}^2(S_y + S_yS_x) + F_{t1u}^1 S_z + F_{t1u}^3(-\frac{3S_z}{5} + Sz^3) & F_{a1g}^0 + 2F_{eg}^2(S_y^2 - \frac{1}{3}S^2) & -F_{t1u}^1 S_x - F_{t1u}^3(-\frac{3S_x}{5} + S_x^3) + F_{t2g}^2(S_yS_z + S_zS_y) \\ -F_{t1u}^1 S_y - F_{t1u}^3(-\frac{3S_y}{5} + S_y^3) + F_{t2g}^2(S_xS_z + S_zS_x) & F_{t1u}^1 S_x + F_{t1u}^3(-\frac{3S_x}{5} + S_x^3) + F_{t2g}^2(S_yS_z + S_zS_y) & F_{a1g}^0 + 2F_{eg}^2(S_z^2 - \frac{1}{3}S^2) \end{pmatrix} \tag{5}$$

**Acknowledgments**

We acknowledge the staff of beamline I21 of Diamond Light Source for their help in setting up and running the experiments. R.-P. Wang and F. Frati are thanked for the discussions.

**Supplementary information for:**
**Novel magnetic excitations beyond the single- and double-magnons**

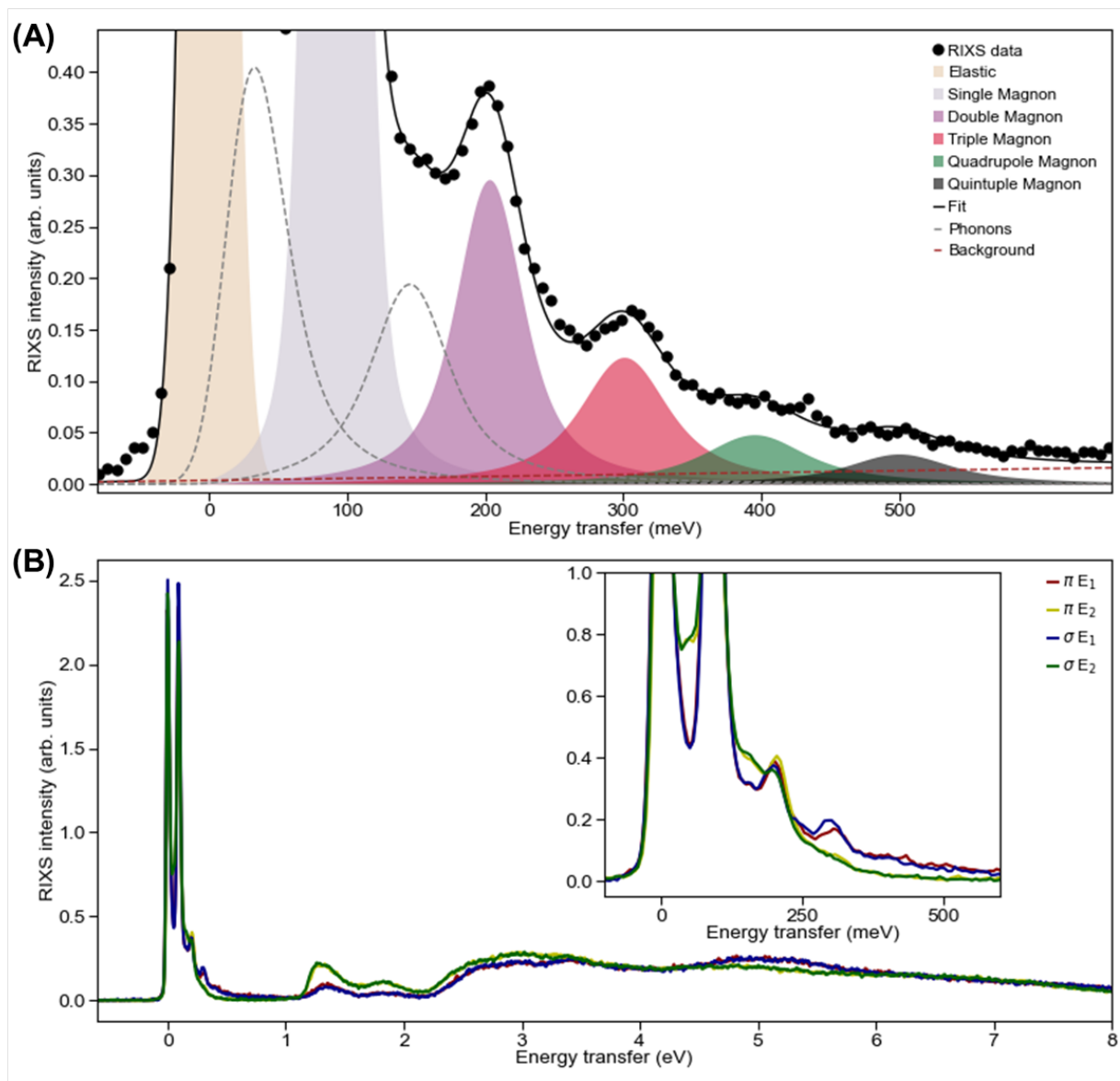

**Fig. S1.**
Panel A. Low energy transfer RIXS spectrum at $E_1$ of XAS, π polarization and α=95°, highlighting the fitted peak profiles of the double- (purple), triple- (red), quadrupole- (green) and quintuple- (black) magnon excitations as described in Methods. Panel B. RIXS spectra showing high energy transfer features at $E_1$ and $E_2$ of XAS for π and σ polarization at α=95°. Inset shows the low energy transfer features. Features close to 43 meV and 150 meV are clearly visible for incident energy of $E_2$.

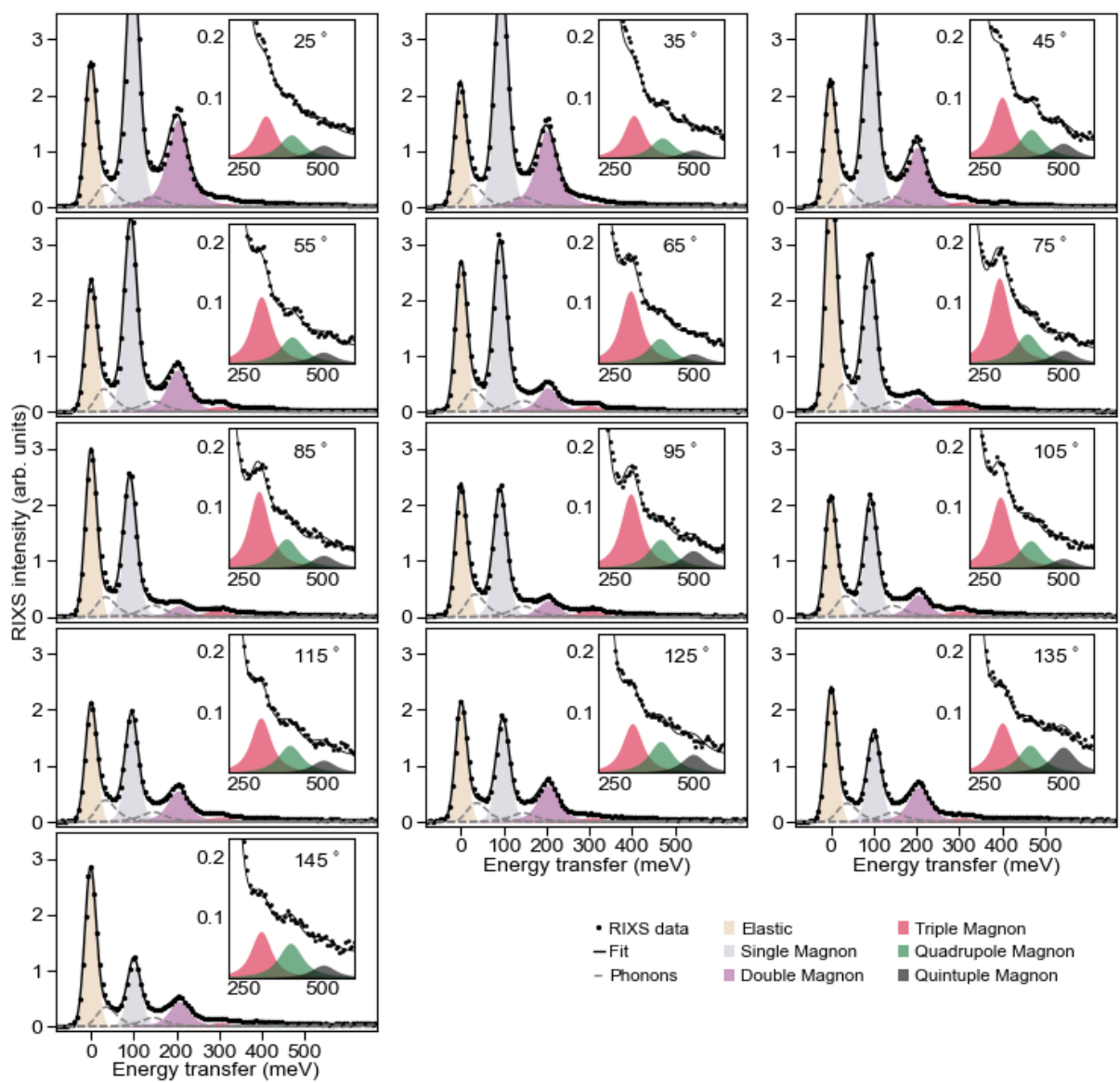

**Fig. S2.**
Low energy RIXS spectra at $E_1$ of XAS along with fitted profiles for π polarization and varying α. The insets show the triple- (red), quadrupole- (green) and quintuple- (black) magnon excitations.

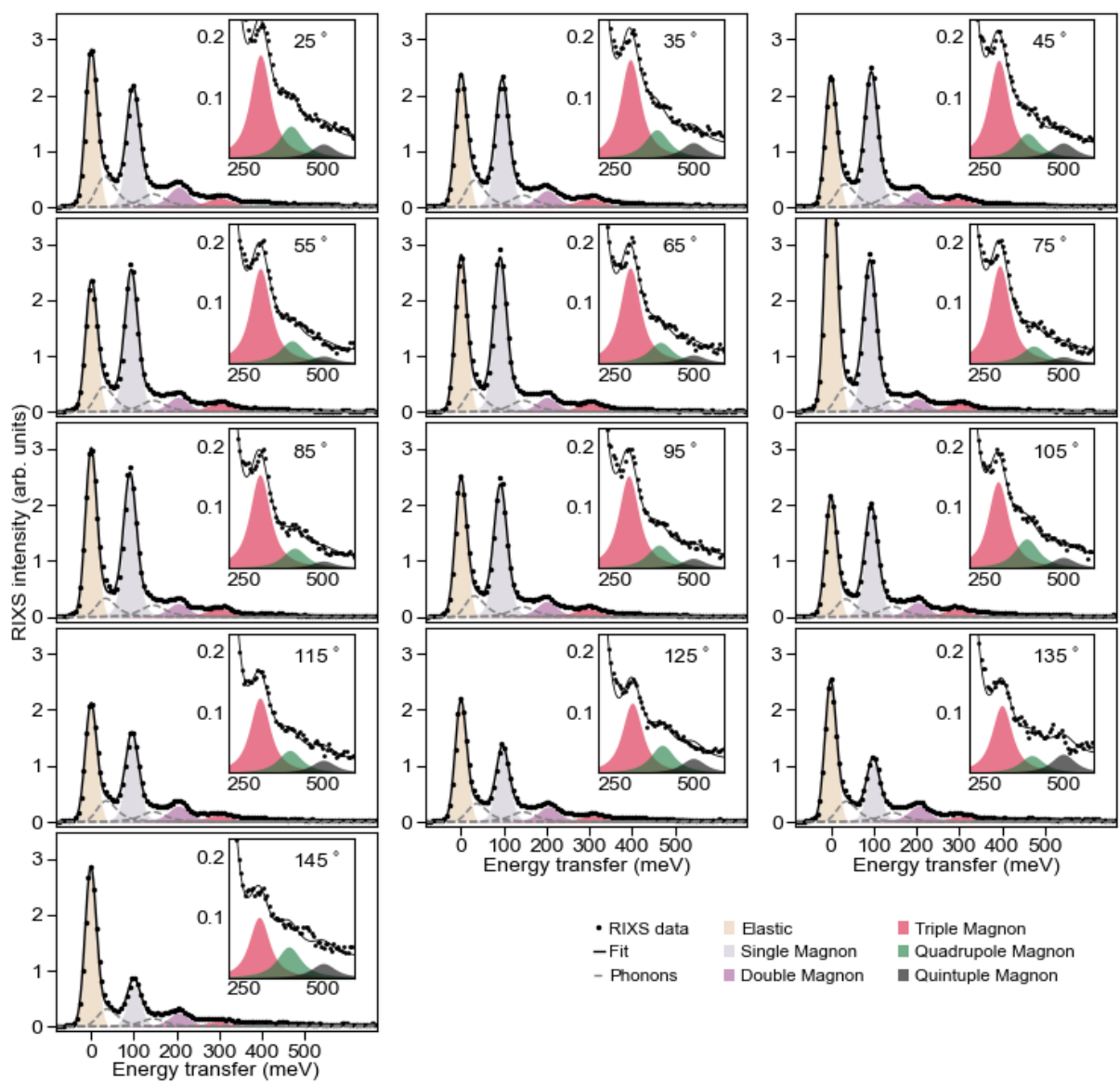

**Fig. S3.**
Low energy RIXS spectra at $E_1$ of XAS along with fitted profiles for σ polarization and varying α. The insets show the triple- (red), quadrupole- (green) and quintuple- (black) magnon excitations.

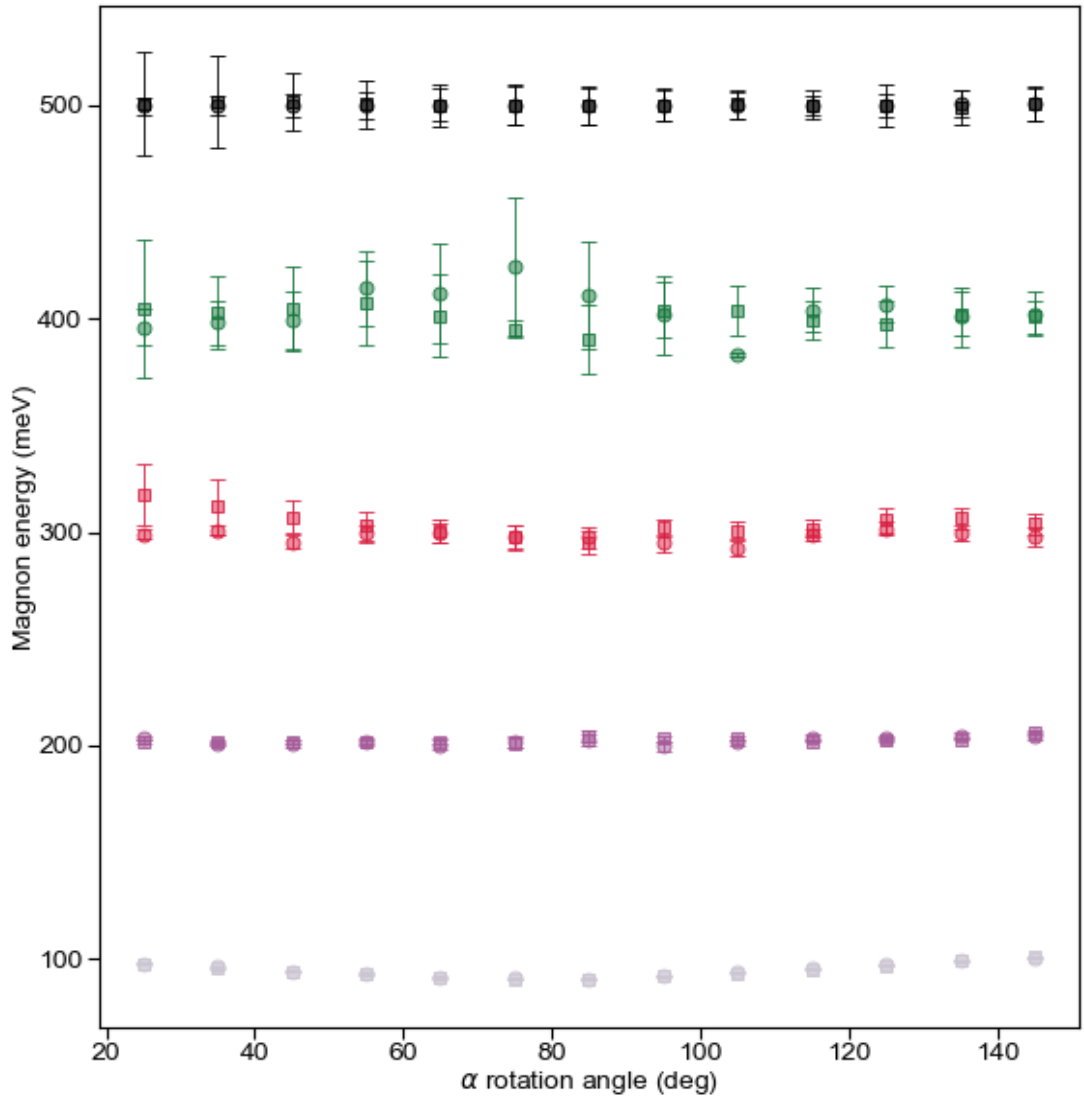

**Fig. S4.**
The energies of single-, double-, triple-, quadrupole- and quintuple-magnon excitations extracted from fitting of RIXS data at $E_1$ of RIXS data for π (squares) and σ (circles) polarisation as described in Methods. The error bars of quintuple-magnon excitations shown are least square fitted energy position errors of the triple magnon)*(5/3).

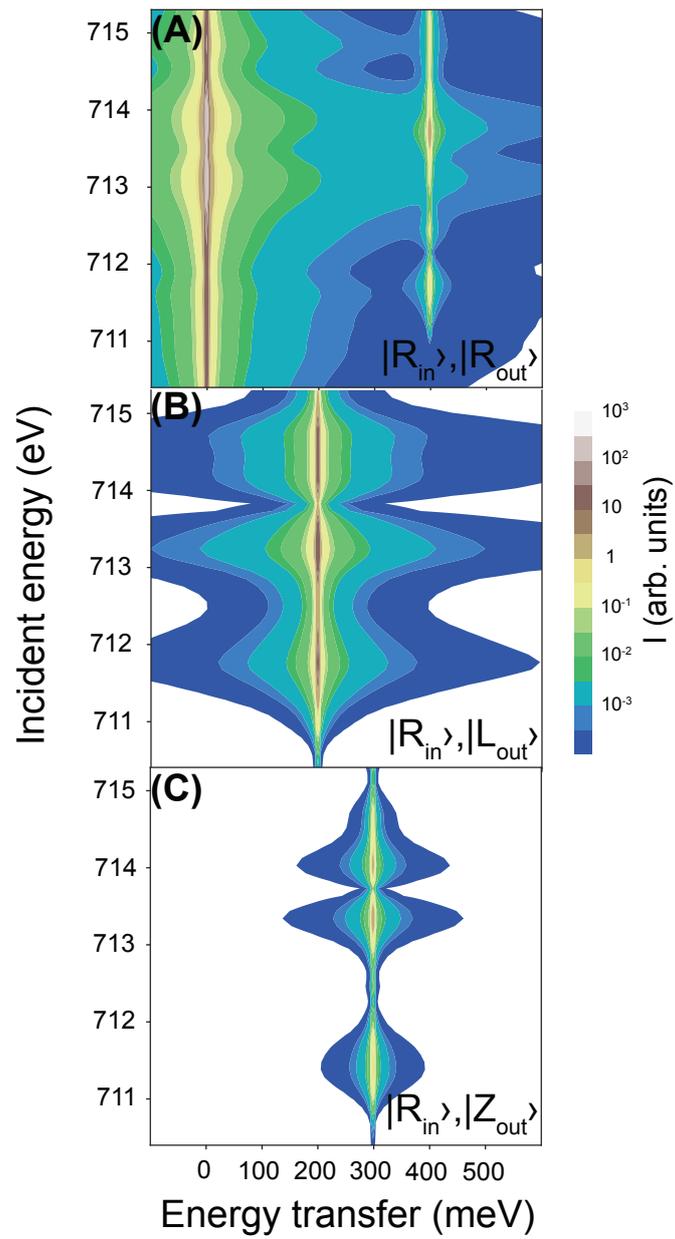

**Fig. S5.**
$Fe^{3+}$ *2p3d* RIXS calculations performed for circular right polarized incoming beam. The outgoing beam polarization is: (A) circular right, (B) circular left, and (C) Z-polarized.

| Parameter | Initials state (eV) | Final state (eV) |
|---|---|---|
| $F_{dd}^{(2)}$ | 8.43 | 8.97 |
| $F_{dd}^{(4)}$ | 5.27 | 5.62 |
| $F_{pd}^{(2)}$ |  | 5.96 |
| $G_{pd}^{(1)}$ |  | 4.45 |
| $G_{pd}^{(3)}$ |  | 2.53 |
| $SOC_d$ | 0.059 | 0.074 |
| $SOC_p$ |  | 8.2 |
| 10Dq | 1.5 | 1.5 |
| Jexch | 0.1 | 0.1 |

**Table S1.**
Parameters used for the *2p3d* RIXS calculation of $Fe^{3+}$ in α-$Fe_2O_3$.